\documentstyle[aps,multicol]{revtex}

\renewcommand{\narrowtext}{\begin{multicols}{2} \global\columnwidth20.5pc}
\renewcommand{\widetext}{\end{multicols} \global\columnwidth42.5pc}
\def\inseps#1#2{\def\epsfsize##1##2{#2##1} \centerline{\epsfbox{#1}}}

\input{epsf.tex}
\begin{document}
\def\ba{\begin{eqnarray}}
\def\ea{\end{eqnarray}}
\def\be{\begin{equation}}
\def\ee{\end{equation}}
\def\tr{{\rm tr}}
\def\gsim{ \lower .75ex \hbox{$\sim$} \llap{\raise .27ex \hbox{$>$}} }
\def\lsim{ \lower .75ex \hbox{$\sim$} \llap{\raise .27ex \hbox{$<$}} }

\title{Constraining Isocurvature Perturbations with CMB Polarisation}

\author{Martin Bucher, Kavilan Moodley and Neil Turok}

\address{DAMTP, Centre for Mathematical Sciences, University of Cambridge\\
Wilberforce Road, Cambridge CB3 0WA, United Kingdom}

\maketitle

\begin{abstract}
The role of cosmic microwave background (CMB) polarisation data
in constraining the presence of primordial isocurvature modes is
examined. While the MAP satellite mission 
will be unable to simultaneously constrain
isocurvature modes and cosmological parameters,
the PLANCK mission will be able to set strong limits on the
presence of isocurvature modes if it makes a precise measurement of
the CMB polarisation sky. We find that if we allow
for the possible presence of isocurvature modes, 
the recently obtained BOOMERANG measurement of the 
curvature of 
the universe fails.
However, a comparably 
sensitive polarisation measurement on the same angular
scales will permit a determination of the curvature of the 
universe without the prior assumption of adiabaticity.   
\end{abstract}
\narrowtext

Measurements of the cosmic microwave background anisotropy
may soon allow us to rigorously determine 
the fundamental character of the primordial
cosmological perturbations with a minimum of simplifying
hypotheses. Recent data \cite{boom,maxima}
already impose formidable constraints 
on the parameter space of simple inflationary
models, and with forthcoming 
satellite missions \cite{mapp,plnck} this 
situation is expected to improve substantially. 

This paper focuses on how to test the assumption of 
adiabaticity, namely that 
all components contributing to the density of the Universe 
are present in spatially uniform ratios
on hypersurfaces of equal cosmic temperature and 
initially share a common velocity field. 
The hypothesis of adiabaticity, put forth initially
on the basis of simplicity, gained support
when it was realized that this was the prediction of the
simplest one-field inflationary iodels\cite{mukh}.
It should be noted, however, that multi-field inflationary
models generically excite isocurvature modes as well
\cite{non-adiab}.

In order to test the hypothesis of adiabaticity through
observation it is necessary to study models where the
primordial perturbations are not solely adiabatic and thus
to attempt to set bounds on the allowed
admixtures of non-adiabatic modes.
Non-adiabatic, or isocurvature, perturbations have already been
studied in the literature, but most 
work has focused on evaluating the viability of cosmological 
models in which the perturbations where entirely isocurvature in 
character, with no adiabatic component at all \cite{old-iso-constr}. 
Observational constraints on an uncorrelated admixture of adiabatic
and cold dark matter isocurvature 
perturbations have also been considered\cite{new-iso-constr}.

Baryon isocurvature (BI)
and cold dark matter isocurvature (CDMI) models were studied some
time ago\cite{old-iso-constr}. More recently it was 
realized that two additional isocurvature
modes are possible: a neutrino isocurvature density (NID) mode and a
neutrino isocurvature velocity (NIV) mode \cite{BMTa,ni2}.
In the NID
mode, the neutrino-photon ratio varies 
spatially. As modes enter the horizon, the photon-baryon fluid begins to
oscillate acoustically whereas 
the neutrinos 
free stream. 
This differential behavior perturbs the 
total energy density, leading to 
structure formation via gravitational clustering.
The neutrino velocity mode assumes 
a relative velocity between the photon and
neutrino components but zero initial
total momentum density. As with the NID
mode, differential evolution spoils this
cancellation after horizon crossing, again leading to structure formation.

In a universe composed of just
photons, baryons, neutrinos, baryons, and a cold dark matter
component, these four isocurvature modes and the adiabatic mode
exhaust the possible modes nonsingular in the $t\to 0$ limit \cite{BMTa}.  
The most general Gaussian primordial perturbation in such a
cosmology is completely characterized by the matrix valued
generalization of the power spectrum
$$ \langle A_a({\bf k})~A_b({\bf k}')\rangle =P_{ab}(k)\cdot
\delta ^3({\bf k}-{\bf k}') $$
where the indices $(a,b=0,1,2,3,4)$ label the modes. When expectation
values of observables quadratic in the linearized perturbations
(such as the CMB $C_l$'s) are considered, the assumption of Gaussianity
is superfluous.  A detailed discussion of how the CMB satellite missions
will be able to constrain the presence of these modes is given 
in ref. \cite{BMTb}. 

We briefly remark on the
possible microphysical origin of these modes. The neutrino isocurvature
mode is rapidly damped before neutrino decoupling, and therefore can
be plausibly produced only by physics operating after $\sim 1 $ second.
The neutrino isocurvature density mode is likewise damped by electroweak
$B+L$ violating anomalous processes (which would convert it mostly into
a baryon isocurvature mode) operating at times earlier than 
$10^{-10}$ seconds. Since both times are well before
photon decoupling ($\sim 10^{13}$ seconds), we nevertheless 
think it legitimate to 
describe the perturbations as `primordial'.

The temperature anisotropy spectra associated with
the regular perturbation modes and their cross-correlations
are shown in Fig.~1. An appropriate power law auto-correlation spectrum was
assumed for each mode, so that the large scale CMB anisotropy is 
approximately scale invariant.
The cross-correlation power spectra were then taken to be 
proportional to the 
geometric mean of the two auto-correlation spectra. Clearly there is
scope for significant generalisation of these assumptions.

\widetext
\begin{figure}
\inseps{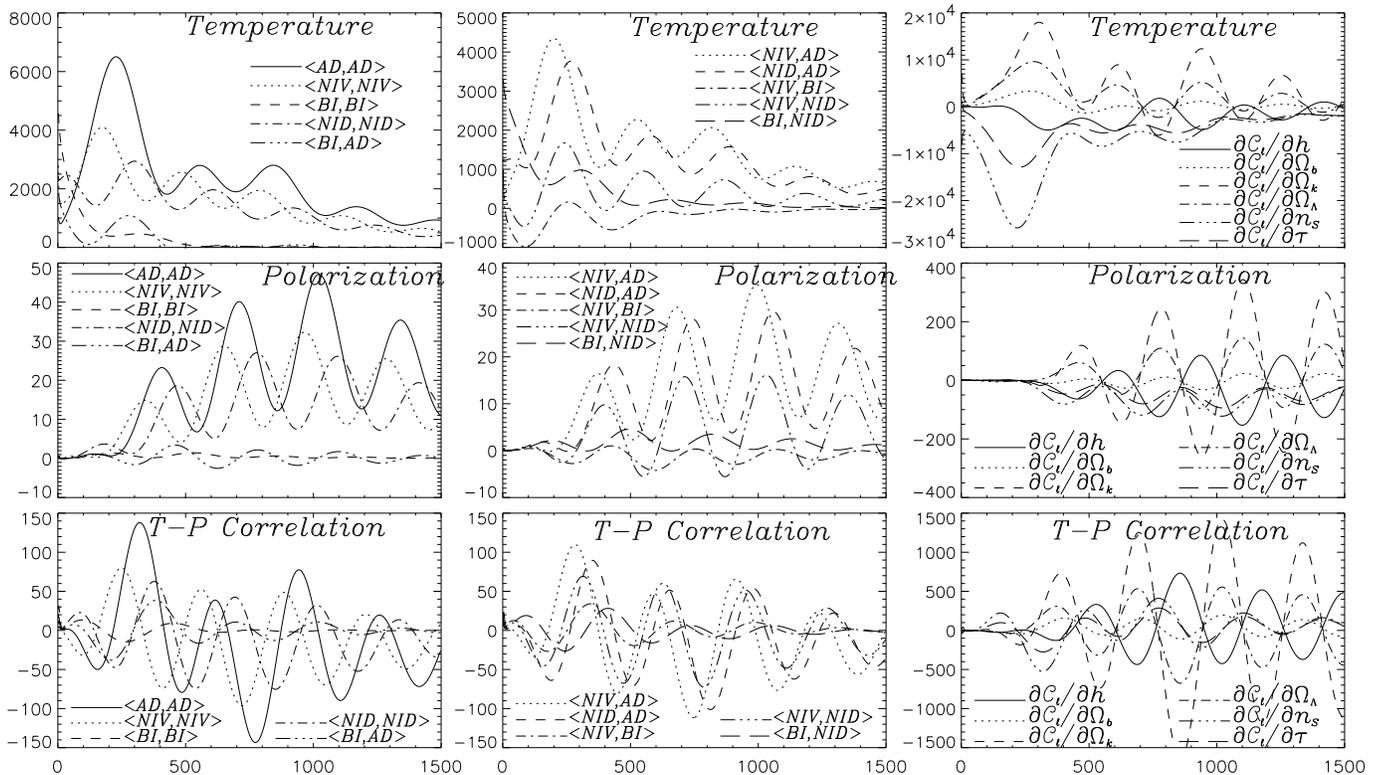}{0.65}
\caption{{CMB multipole spectra for the various modes, their
cross-correlations, variations in the cosmological parameters.}
From top to bottom the rows show $l(l+1)C_l/2\pi $ for 
the temperature, polarization, and temperature-polarization cross 
correlation, respectively, in $\mu K.$ The
$C_l$ spectra for the various modes and their cross 
correlations are shown in the first two columns. The 
rightmost column shows the derivatives of the 
spectra with respect to the different 
cosmological parameters. The modes are indicated as follows:
adiabatic (AD), neutrino isocurvature velocity (NIV), 
baryon isocurvature (BI), and neutrino isocurvature density (NID).
A fiducial model with the parameter choices 
$\Omega_b=0.06,~\Omega_\Lambda=0.69,
~\Omega_{cdm}=0.25,~h=0.65,~\tau_{reion} =0.1$ and $n_s=1$ has been 
assumed. Because the  CDM isocurvature mode produces a spectrum 
nearly identical to that of the BI mode, 
it is not considered separately.}
%\vskip .1in
\label{fig:multi}
\end{figure}
\narrowtext

The adiabatic $C_l$ temperature spectrum is characterised by 
a flat Sachs-Wolfe plateau at low $l$ and a series of acoustic peaks,
at $l\approx 220+310n$, $(n=0,1,\ldots )$ in a flat Universe.
Scale invariant 
baryon and CDM isocurvature
spectra produce little power at high $l$.
The NID mode exhibits 
the phase shift characteristic 
of isocurvature density modes but 
the NIV mode produces a pattern of peaks 
more similar to that of the adiabatic mode.
The NIV mode acquires a nearly cancelling phase shift because
the velocity is out of phase with the density. 
Note in particular how similar the 
NIV-adiabatic mode cross correlation $C_l$ is
to the pure adiabatic mode $C_l$.  
The polarisation and temperature-polarisation cross correlation 
power spectra associated with the isocurvature modes are 
indicated in the bottom two rows of Fig.~1.

How feasible will it be to
constrain or to detect isocurvature modes using CMB measurements? 
The key question is whether one can distinguish the effect
of the isocurvature modes from 
those of variations in the cosmological parameters (see e.g. \cite{selj}).
The derivatives of the $C_l$ power spectra about 
a fiducial $\Lambda$CDM model with respect
to cosmological parameters are shown 
in the rightmost column of Fig.~1.
For small admixtures of isocurvature modes, 
the question is whether linear combinations of these spectra
can be distinguished from those of the first two columns of Fig.~1.

For small variations of cosmological parameters 
and small admixtures  of isocurvature modes
one can parameterize the likelihood function as a multivariate 
Gaussian about a fiducial adiabatic model. Using projected 
estimates of the instrument noise for the MAP and PLANCK
satellite missions, we have computed the estimated errors
on the cosmological parameters and isocurvature auto-
and cross-correlation amplitudes assuming the sky 
is actually described by a simple adiabatic $\Lambda $CDM
model. Table I shows that
the MAP satellite will be unable to simultaneously
constrain isocurvature modes and measure the cosmological
parameters. The PLANCK satellite will, but only if it 
measures the CMB sky polarisation as accurately as currently
planned.

We now discuss precisely how the polarisation measurement
resolves the degeneracy between isocurvature modes and
cosmological parameters. In Fig.~2 we illustrate how
polarization serves to remove the degeneracy corresponding to the
eigenvector pointing in the most
uncertain (i.e., flattest) direction of the relative likelihood when only
temperature information is taken into account.

\widetext
\begin{table}
%\centering
\begin{tabular}{|| c | c | c| c | c | c | c | c | c | c ||}
 & MAP & MAP & MAP & MAP & PLANCK & PLANCK & PLANCK & PLANCK & PLANCK\cr
& T & TP & T & TP & T & TP & T & T+P & TP\cr
& adia & adia & all & all & adia & adia & all & all & all\cr
& only & only & modes & modes & only & only & modes & modes & modes\cr
\hline
$\delta h/h$ &      12.37 &      7.42 &      175.84 &      20.40 &
      9.93 &      3.69 &      40.13 &      7.31 &      4.36\cr
\hline
$\delta \Omega _b/\Omega _b$ &      27.76 &      13.34 &      325.38 &
      28.57 &      19.37 &      7.26 &      68.85 &      14.42 &
      8.61\cr
\hline
$\delta \Omega _k$ &      9.79 &      2.72 &      75.32 &      4.55
 &      4.92 &      1.83 &      20.56 &      3.59 &      2.18\cr
\hline
$\delta \Omega _\Lambda /\Omega _\Lambda $ &      12.92 &      5.02 &
      123.63 &      18.53 &      2.74 &      1.21 &      5.93 &
      2.45 &      1.49\cr
\hline
$\delta n_s/n_s$ &      7.02 &      1.62 &      89.89 &      6.53 &
     0.73 &     0.37 &      3.92 &     0.90 &     0.70\cr
\hline
$\tau _{reion}$ &      37.39 &      1.81 &      104.81 &      2.23 &
      8.25 &     0.41 &      35.35 &     0.74 &     0.56\cr
\hline
$\langle NIV,NIV\rangle $&.. &.. &      114.34 &      11.47 &.. &.. &
      43.45 &      1.36 &      1.14\cr
\hline
$\langle BI,BI\rangle $ &.. &.. &      573.46 &      29.71 &.. &.. &
      53.29 &      6.16 &      4.23\cr
\hline
$\langle NID,NID\rangle $ &.. &.. &      351.79 &      29.87 &.. &.. &
      19.18 &      4.77 &      2.37\cr
\hline
$\langle NIV,AD\rangle $ &.. &.. &      434.70 &      44.06 &.. &.. &
      121.59 &      8.21 &      4.69\cr
\hline
$\langle BI,AD\rangle $ &.. &.. &      1035.02 &      59.25 &.. &.. &
      58.75 &      15.03 &      8.97\cr
\hline
$\langle NID,AD\rangle $ &.. &.. &      1287.60 &      67.49 &.. &.. &
      114.39 &      13.87 &      5.77\cr
\hline
$\langle NIV,BI\rangle $ &.. &.. &      601.70 &      32.29 &.. &.. &
      46.91 &      7.72 &      3.67\cr
\hline
$\langle NIV,NID\rangle $ &.. &.. &      744.00 &      46.46 &.. &.. &
      80.01 &      7.55 &      2.97\cr
\hline
$\langle BI,NID\rangle $ &.. &.. &      534.32 &      39.11 &.. &.. &
      100.97 &      7.56 &      4.60\cr
\end{tabular}
\caption{This table indicates the one sigma percentage errors 
on cosmological parameters and
isocurvature mode amplitudes anticipated for the MAP and PLANCK satellite
experiments. In the column headers,
T denotes constraints inferred
from temperature measurements alone, 
TP those from the complete temperature and polarisation measurements,
and 
T+P those inferred if
temperature and polarisation information 
is used separately without including the cross-correlation.
} 
\end{table}
\vskip -.45in
\narrowtext
\vskip -.25in

\begin{figure}
\inseps{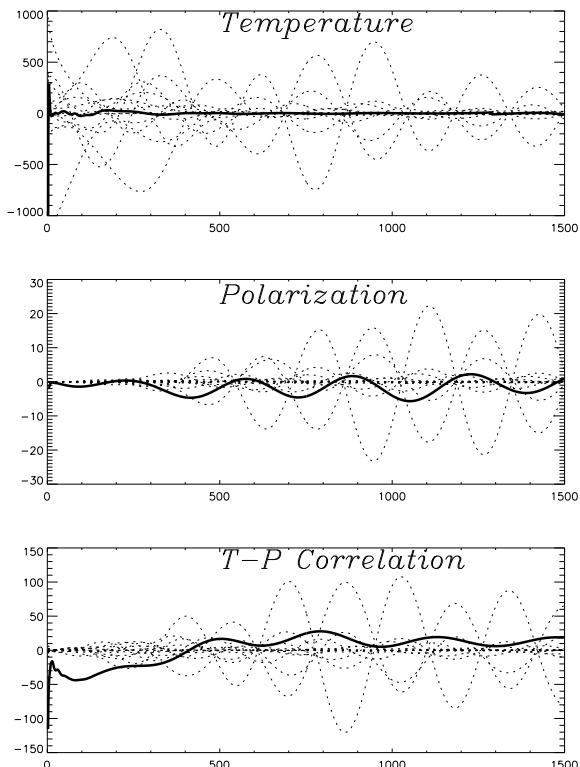}{0.67}
\vskip -.10in
\caption{{\bf Breaking the degeneracies with polarization.}
The top panel indicates the delicate cancellation in the 
temperature power spectrum between the various components of
the most uncertain principal direction. 
The lower panels show how this cancellation is 
broken in the polarization and
temperature-polarization cross-correlation spectra.
}
\label{fig:cancela}
\end{figure}
\begin{figure}
%\inseps{pol_cross_deg_brk.ps}{0.47}
\inseps{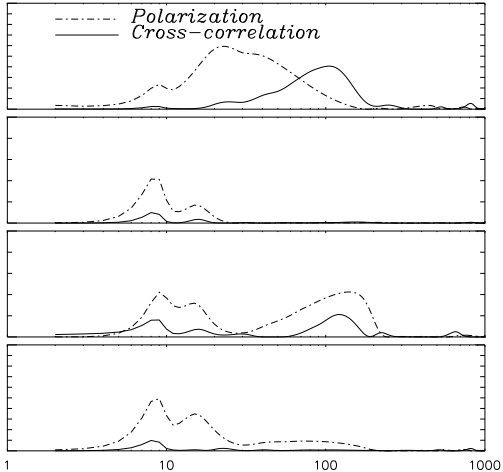}{0.47}
\caption{
When only temperature information from PLANCK is taken into account,
the uncertainties in the four most poorly measured principal
directions are 239\%,  60\%, 36\%, and 23\%. These numbers are
the inverse square root of the corresponding eigenvalues of the
Fisher matrix.
When the
polarization information anticipated from PLANCK is taken into account
as well, these uncertainties are reduced to
11.1\%, 10.3\%, 6.6\%, and 4.6\%, respectively. The
plots (from top to bottom, respectively) indicate the contributions of
the polarisation information at each $l$ to diagonal elements of
the Fisher matrix in
these directions. The cross correlation contribution is by definition
the difference between the total and that obtained from polarisation
and temperature information taken separately.
}
\label{fig:pol-cross-deg-brk}
\end{figure}

\noindent
The dotted curves 
are the contributions of the various components of the eigenvector.
Summing these contributions, one finds very little net 
contribution to the temperature anisotropy (solid line). This sum
is shown multiplied by ten for clarity.
The lower two panels indicate the corresponding curves
for the polarization and temperature-polarization
cross-correlation, respectively. For these the
delicate cancellation is broken, indicating that
the polarization and temperature-polarization
cross-correlation provide the information required
to break the degeneracy.
Fig.~\ref{fig:pol-cross-deg-brk} shows
constraints on the four smallest temperature eigenvalues that will be
provided by a PLANCK measurement of the polarisation and
cross-correlation spectra. The area under the curve
gives the Fisher information, or log likelihood in the
Gaussian approximation, provided by the
polarisation measurements.  The degeneracy breaking by PLANCK polarisation
measurements occurs primarily at 
$l\lsim 100$. Fig.~2 shows that 
there is in fact considerable degeneracy breaking at higher $l$, but
this is not detectable with the instrument noise anticipated 
for PLANCK.

To see why considerable information resides in the range
$l\lsim 100$, note that on these scales one
directly observes the 
primordial (super-horizon scale) polarisation. This is very
different for the adiabatic and isocurvature modes (Fig. ~4).
In particular, for
the NIV mode, since the large scale CMB anisotropy comes mainly from the Doppler
effect rather than the Sachs-Wolfe effect, 
the polarisation spectrum is enhanced       
by a factor of $l^{-2}$ at low $l$ 
relative to that for the adiabatic mode.
The effect at very low $l$ of varying $\tau_{reion}$, is well 
known \cite{selj}.

\begin{figure}
\inseps{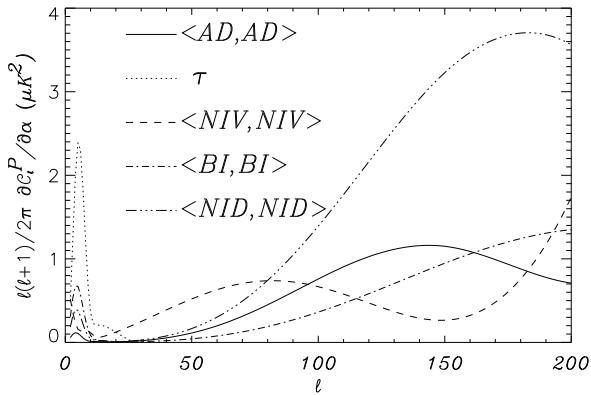}{0.47}
\caption{Polarisation signal of the various modes on large angular scales
($l\lsim 100$). Measurements on these scales are largely responsible for
the degeneracy breaking which polarisation measurements allow.
}
\label{fig:low-l-pol}
\end{figure}
\vskip -.1in
Finally, one can ask to what extent the constraint on the curvature 
of the universe, derived using the recent BOOMERANG data, 
$\delta \Omega_k < 0.12$ at $2\sigma$ \cite{boom},
is affected by the possible presence
of isocurvature modes. Adopting the above flat fiducial model,
we have computed the 
errors in $\delta \Omega_k$ allowing for
sample variance and using the
published values for the
instrument noise \cite{boom}.
We do not account for a calibration uncertainty and fix all 
cosmological parameters except $\Omega_k$. We then allow 
arbitrary amounts of isocurvature and cross-correlation power
and attempt to set limits simultaneously on these and
on $\delta \Omega_k$. With the assumption of adiabaticity, the
data yield a one sigma error on $\delta \Omega_k$ of $2\%$. However,
with isocurvature modes allowed, this error rises to 
$577\%$. Of course, the approximation that the likelihood is
Gaussian breaks down at such a large level. Nevertheless,
the conclusion that the error in $\delta \Omega_k$ is
of order unity is firm. If we assume that it will be possible to 
make a 
BOOMERANG-like polarisation map with accurate source subtraction
and that the polarisation error is optimal, $\sigma_P^2=2\sigma_{T,BOOM}^2$, 
then we find that the additional polarisation information allows the 
constraint 
on $\Omega_k$ to be reduced to $13\%$, a very significant improvement.
This work therefore provides strong
motivation for such a polarisation measurement.

{\bf Acknowledgements:} We thank 
Chris Barnes, Dick Bond, Robert Crittenden, 
David Langlois,  Lyman Page, 
David Polarski, and John Ruhl for useful discussions.
MB acknowledges support from Dennis Avery.
KM acknowledges support from 
the Commonwealth Scholarship Commission in the UK. 
The research of NT is supported by PPARC (UK). 
Computations were carried out using the COSMOS 
supercomputer funded by PPARC.

\widetext
\end{document}